\documentclass[twocolumn]{aastex631}

\shorttitle{Tidal Evolution of a Hypothetical Venus Moon}
\shortauthors{Stephen R. Kane}

\begin{document}

\title{Tidal Demise: The Evolution and Fate of a Hypothetical Venus Moon}

\author[0000-0002-7084-0529]{Stephen R. Kane}
\affiliation{Department of Earth and Planetary Sciences, University of California, Riverside, CA 92521, USA}
\email{skane@ucr.edu}

\author[0000-0001-9619-5356]{Franck Selsis}
\affiliation{Laboratoire d'astrophysique de Bordeaux, University of Bordeaux, CNRS, 33615, Pessac, France}

\author[0000-0002-3555-480X]{J\'er\'emy Leconte}
\affiliation{Laboratoire d'astrophysique de Bordeaux, University of Bordeaux, CNRS, 33615, Pessac, France}

\author[0000-0001-8974-0758]{Sean N. Raymond}
\affiliation{Laboratoire d'astrophysique de Bordeaux, University of Bordeaux, CNRS, 33615, Pessac, France}


\begin{abstract}

Venus possesses no natural satellite, raising the question whether a
formed moon could have survived. We explore the tidal evolution of a
Venus-moon system, coupling Venus's spin to the satellite's orbit
under tides from the moon and Sun. We survey spin period ($P_0 =
5$--100~hr), moon mass ($M_m = 0.01$--$10~M_{\rm Moon}$),
eccentricity, quality factor, and initial semi-major axis under both
constant-$Q$ and constant time lag models. Survival depends on
competition between outward migration ($\propto M_m$) and synchronous
radius expansion ($\propto M_m^2$): for circular orbits around rapidly
spinning Venus ($P_0 \lesssim 12$~hr), a lunar-mass satellite survives
the age of the Solar System in both models. For $P_0 \lesssim 10$~hr,
eccentricity pumping can destabilize low-mass satellites, while for
$P_0 \gtrsim 15$~hr or $M_m \gtrsim 2~M_{\rm Moon}$ the synchronous
radius overtakes the orbit and drives Roche destruction within
$\sim$0.03--1.7~Gyr in the constant-$Q$ model. The constant time lag
model instead permits quasi-synchronous survival for massive moons at
fast spin. Explaining Venus's present state requires satisfying two
constraints simultaneously: loss of the satellite and despinning of an
initially rapid rotator. Both are met only within a restricted region
of parameter space, favoring moderate post-impact spin periods and
lunar-to-super-lunar masses.  Giant impact simulations predict spin
periods $\gtrsim$12~hr for Venus's present rotation, placing a
lunar-mass satellite at the survival boundary. For last-impact
conditions within this region, the present absence of a Venusian
satellite arises through tidal evolution alone; a subsequent
catastrophic stripping event, while capable of removing a moon, is not
required.

\end{abstract}

\keywords{astrobiology -- planetary systems -- planets and satellites:
  dynamical evolution and stability -- planets and satellites:
  individual (Venus)}


\section{Introduction}
\label{sec:intro}

The tidal evolution of planetary satellites is among the most
consequential dynamical processes in the Solar System, governing the
orbital histories of moons, the rotational states of their host
planets, and the long-term habitability of terrestrial worlds
\citep{goldreich1966a,murray1999a,barnes2002b}. The Earth-Moon system
provides the archetypal example: angular momentum transfer from
Earth's rapid rotation has driven the Moon outward from an initial
distance of $\sim$3.5~$R_\oplus$ to its present semi-major axis of
60~$R_\oplus$ over 4.5~Gyr
\citep{touma1994b,canup2001a,cuk2016b,farhat2022b}. This process has profoundly influenced
Earth's obliquity stability, tidal dissipation budget, and climate
evolution
\citep{laskar1993b,lissauer2012a,green2019}.

Venus presents a striking counterpoint to Earth. Despite their nearly
identical masses and radii, Venus today rotates retrograde with a
period of 243~days and possesses no natural satellite. The absence of
a Venusian moon has long been noted as one of the fascinating puzzles
of comparative planetology, with implications for the divergent
evolutionary pathways of Earth and Venus
\citep{kane2014e,way2020,kane2024b}. Several mechanisms have been
proposed to explain this absence, broadly falling into three
categories: Venus never acquired a moon through the accretion process,
a moon formed but was tidally destroyed, or a moon formed but was
removed through a subsequent giant impact. The first category
encompasses the possibility that Venus never experienced a
moon-forming giant impact at all. Several recent formation models
suggest that Venus may have grown largely through pebble accretion or
torque-driven embryo migration during the disk phase, avoiding the
late giant impacts that characterize some Earth formation scenarios
\citep{johansen2021,broz2021c,nesvorny2025b,huang2025c}. The present
work instead addresses the second category, asking whether a moon, if
formed through a giant impact, could survive to the present day.

Late-stage accretion models predict that Venus likely experienced at
least one giant impact capable of producing a circumplanetary debris
disk \citep{canup2001a}, raising the possibility that a prograde moon
could have been subsequently destroyed through spin reversal caused by
later impacts. \citet{bills1992a} provided an early analysis of this
scenario, demonstrating that any Venus satellite would undergo orbital
decay, form an ephemeral ring upon crossing the Roche limit, and
ultimately produce surface cratering upon reaccretion. An important
corollary is that reaccretion of a moon onto Venus's surface must have
occurred early, while the planet was still molten or partially molten,
or through an event energetic enough to remelt the surface. Late
reaccretion would imprint a large-scale surface signature that is not
observed. \citet{jacobson2017} further argued that Venus's lack of
both a moon and an internally-generated magnetic field may reflect the
preservation of primordial compositional stratification, a consequence
of Venus having accreted through many smaller impacts rather than a
small number of energetic giant impacts. More recently,
\citet{makarov2024a} explored the capture of a retrograde satellite
(dubbed ``Neith'') through chaotic interactions with a remnant debris
disk, finding that such a moon would brake Venus's prograde rotation,
eventually reverse it, and spiral inward to destruction at the Roche
limit. \citet{makarov2025d} extended this analysis using
rheology-independent synchronization conditions, showing that the
synchronization criterion is never satisfied for a retrograde moon of
Venus, making inward spiral inevitable. \citet{bussmann2025} performed
SPH simulations of giant impacts on Venus and found that, for impact
geometries consistent with Venus's present-day rotation, the resulting
debris would likely reaccrete onto the planet, preventing the
formation of long-lasting satellites.

Previous work has discussed the ease with which an inner-planet
satellite can be lost. \citet{ward1973a} and \citet{burns1973b}
demonstrated that solar tidal perturbations, acting on a moon whose
host planet is being tidally despun, can drive the satellite either
outward to escape or inward to destruction on geologically short
timescales. \citet{burns1973b} claimed that the present absence of
satellites at Mercury and Venus is not compelling evidence that such
satellites never existed, since their removal is a natural dynamical
outcome. The broader context is the frequency and consequence of giant
impacts during terrestrial planet formation. At least one such impact
is firmly established for Earth, and the formation of Mercury's large
core is often attributed to a similar event; some scenarios invoke
multiple impacts to satisfy the Earth-Moon angular momentum budget
\citep{cuk2012c}. However, not all accretion histories are
impact-dominated. Geochemical evidence such as Mercury's high Cl/K
ratio \citep{evans2015b} and the inferred preservation of primordial
compositional stratification within Venus \citep{jacobson2017} has
been interpreted as evidence for relatively quiescent accretion
through many smaller impacts. Whether every giant impact produces a
long-lived moon, and what becomes of those moons, remains an open
question that motivates the present study.

These studies establish that the dynamical history of a hypothetical
Venus moon is fundamentally shaped by the interplay between Venus's
rotation rate, the tidal quality factor of Venus's interior, and the
mass of the satellite. However, a systematic exploration of this
parameter space using the complete tidal evolution equations has not
been performed. The tidal framework of \citet{hut1981c}, as extended
by \citet{leconte2010a} to arbitrary eccentricity and obliquity,
provides exact solutions in the viscous (constant time lag; CTL)
approximation and has been widely applied to exoplanetary systems
\citep{levrard2007a,bolmont2015}. The complementary constant-$Q$
framework of \citet{goldreich1966a}, as applied to satellite stability
by \citet{barnes2002b}, yields analytical timescale estimates that
provide useful physical intuition. The two models make qualitatively
different predictions near the synchronous radius, where the
constant-$Q$ torque changes sign discontinuously while the CTL torque
passes smoothly through zero.

The fate of a Venus moon carries direct implications for early Venus
habitability. A large satellite would stabilize obliquity
\citep{laskar1993b}, drive ocean tides that influence rotational
evolution \citep{green2019}, and potentially power a geodynamo through
tidal heating \citep{stevenson2003}. The Venus Zone framework
\citep{kane2014e,ostberg2023a} and the concept of Venus as an anchor
point for planetary habitability \citep{kane2024b} both emphasize the
importance of understanding why Venus diverged from Earth. The
presence or absence of a moon during Venus's early history may
represent a critical branch point in this divergence. Moreover,
slowly-rotating terrestrial exoplanets in the Venus Zone should
generically lose their moons on shorter timescales than the system
age, with implications for obliquity stability and habitability
assessments of these worlds.

In this paper, we construct a semi-analytical framework for the
coupled spin-orbit evolution of a hypothetical Venus-moon system and
survey the parameter space governing the moon's survival. In
Section~\ref{sec:framework}, we describe the tidal models employed and
their relationship to Venus's physical properties. In
Section~\ref{sec:results}, we present the results of the parameter space
exploration under both the constant-$Q$ and CTL models, including the
effects of initial spin period, moon mass, tidal quality factor,
initial semi-major axis, and orbital eccentricity. In
Section~\ref{sec:disc}, we discuss the physical interpretation of
the results, the comparison between tidal models, and the
implications for early Venus habitability and exoplanetary science.
Section~\ref{sec:con} summarizes the conclusions and suggests
directions for future work.


\section{Analytical Framework}
\label{sec:framework}

Here we describe the tidal models and system parameters used in our
calculations.


\subsection{System Parameters}
\label{sec:params}

We adopt the following physical parameters for Venus: mass $M_V =
4.8675 \times 10^{24}$~kg, radius $R_V = 6051.8$~km, orbital
semi-major axis $a_V = 0.7233$~AU, orbital period $P_V = 224.7$~days,
tidal Love number $k_{2,V} = 0.295$ \citep{konopliv1996}, and moment
of inertia factor $C/(M_V R_V^2) = 0.337$ \citep{margot2021c}. We note
that the tidal response of Venus is poorly constrained observationally
due to the lack of Venus satellite whose orbital evolution could be
measured, resulting in a range of quoted Love numbers (e.g., $k_2 =
0.25$; \citealt{murray1999a}). Because our calculations parameterize
the dissipation through the quality factor $Q_V$, which we survey
across more than an order of magnitude, the results are insensitive to
the precise adopted value of $k_{2,V}$. The Roche limit for a fluid
satellite with lunar density ($\rho_m = 3344$~kg~m$^{-3}$) orbiting
Venus ($\rho_V = 5243$~kg~m$^{-3}$) is $a_{\rm Roche} = 2.85~R_V$. The
Hill sphere radius is $R_H = 167~R_V$, and the critical semi-major
axis for prograde satellite stability is $a_{\rm crit} = 0.49~R_H =
82~R_V$ \citep{holman1999,domingos2006}. Throughout this work we
consider a prograde satellite, orbiting in the same sense as Venus's
(initial) rotation; the retrograde case, in which solar tides and
satellite tides act in concert to drive inevitable inward spiral, has
been treated separately by \citet{makarov2024a} and
\citet{makarov2025d}.

The synchronous radius, the orbital distance where the satellite's
mean motion equals Venus's spin rate, is given by $a_{\rm sync} = (G
M_V / \Omega_V^2)^{1/3}$, where $\Omega_V$ is Venus's angular
rotation rate. This radius sets the boundary between outward migration
(moon outside $a_{\rm sync}$) and inward migration (moon inside
$a_{\rm sync}$). For reference, $a_{\rm sync} = 3.6~R_V$ for a
10-hour spin period, $6.5~R_V$ for 24 hours, and $10.3~R_V$ for 48
hours. At Venus's present-day rotation period of 243~days, $a_{\rm
sync} \approx 254~R_V$, which significantly exceeds the Hill radius of $167~R_V$. This means that
no stable synchronous orbit currently exists for Venus, and any satellite at a
dynamically stable distance would experience inward migration
toward the Roche limit.


\subsection{Constant-$Q$ Model}
\label{sec:cqmodel}

In the constant-$Q$ framework
\citep{goldreich1966a,murray1999a,barnes2002b}, the tidal torque on
Venus due to the moon is
\begin{equation}
  \tau_{\rm m} = -\frac{3}{2} \frac{k_{2,V} G M_m^2 R_V^5}{Q_V a_m^6}
  \, {\rm sgn}(\Omega_V - n_m)
  \label{eq:torque_cq}
\end{equation}
where $M_m$ is the satellite mass, $a_m$ is the satellite semi-major
axis, $n_m = \sqrt{G M_V / a_m^3}$ is the orbital mean motion, and
$Q_V$ is the tidal quality factor. The solar tidal torque on Venus
follows the same form with the substitutions $M_m \rightarrow
M_\odot$ and $a_m \rightarrow a_V$
\citep{barnes2002b}. The coupled evolution equations for Venus's spin
rate and the satellite's semi-major axis are
\begin{equation}
  \frac{d\Omega_V}{dt} = \frac{\tau_{\rm m} + \tau_\odot}{C_V}
  \label{eq:spin_cq}
\end{equation}
\begin{equation}
  \frac{da_m}{dt} = -\frac{2\sqrt{a_m} \, \tau_{\rm m}}{M_m
    \sqrt{GM_V}}
  \label{eq:orbit_cq}
\end{equation}
where $C_V$ is Venus's moment of inertia. Equation~\ref{eq:orbit_cq}
is equivalent to the standard constant-$Q$ orbital evolution
expression given by \citet[Equation~4.213]{murray1999a}. The
characteristic satellite lifetime follows from \citet{barnes2002b}
Equation~7:
\begin{equation}
  T = \frac{2}{13} \frac{a_{\rm crit}^{13/2} \, Q_V}{3 k_{2,V} M_m
    R_V^5 \sqrt{M_V/G}}
  \label{eq:lifetime}
\end{equation}
which provides an upper bound on the survival time.


\subsection{Constant Time Lag Model}
\label{sec:ctlmodel}

In the CTL framework \citep{hut1981c}, the tidal response is
characterized by a fixed time delay $\Delta t$ between the tidal
potential and the resulting deformation. The torque on Venus due to
the satellite is proportional to $(\Omega_V - n_m)$ rather than ${\rm
  sgn}(\Omega_V - n_m)$:
\begin{equation}
  \tau_{\rm m}^{\rm CTL} = -3 k_{2,V} \Delta t \frac{G M_m^2
    R_V^5}{a_m^6} (\Omega_V - n_m)
  \label{eq:torque_ctl}
\end{equation}
and the orbital evolution is
\begin{equation}
  \frac{da_m}{dt} = \frac{6 k_{2,V} \Delta t \sqrt{G} \, M_m R_V^5}
       {\sqrt{M_V} \, a_m^{11/2}} (\Omega_V - n_m)
       \label{eq:orbit_ctl}
\end{equation}
The primary distinction from Equation~\ref{eq:torque_cq} is that the
CTL torque vanishes continuously at $\Omega_V = n_m$, admitting a
tidal fixed point at the synchronous configuration. Following the
convention of \citet{leconte2010a}, we calibrate the CTL model through
the modified quality factor $Q'$, such that $k_2/Q' = k_2 \Delta t
\sigma$ at the chosen reference tidal frequency $\sigma$ (their
Equation~10). This prescription is used to match the CTL and
constant-$Q$ models at the fiducial configuration; away from that
frequency, the effective dissipation is frequency-dependent:
high-frequency tides (far from synchronization) are more strongly
damped, while tides near the synchronous frequency are only weakly
dissipated.

For eccentric orbits, we employ the full Hut equations as derived in
\citet{leconte2010a}, which are exact in the viscous approximation for
arbitrary eccentricity. The semi-major axis and eccentricity evolution
equations involve the eccentricity functions $N(e)$, $N_a(e)$,
$\Omega(e)$, $\Omega_e(e)$, and $N_e(e)$, as defined in Equations~2--8
of \citet{leconte2010a}. These functions contain factors as divergent
as $(1 - e^2)^{-15/2}$ for $e \gtrsim 0.3$, leading to qualitatively
different evolution histories compared with
the commonly used $e^2$-truncated approximations. The spin evolution
equation (Equation~9 of \citealt{leconte2010a}) and the tidal
dissipation rate at pseudo-synchronization (their Equation~13) are
similarly employed.


\subsection{Parameter Space}
\label{sec:paramspace}

We survey the following parameter space. For the initial Venus spin
period, we consider $P_0 = 5$--100~hr, spanning the range from a
rapidly spinning post-impact Venus to a moderately slow rotator. A
spin period of 5~hr corresponds to the canonical Moon-forming impact
scenario; faster initial rotation is entirely plausible within our
understanding of terrestrial planet formation, and high angular
momentum impact models can leave the planet spinning with periods as
short as $\sim$2~hr \citep{cuk2012c}. Our lower bound is therefore
conservative, and the qualitative behavior we identify at fast spin
extends to even shorter periods. The satellite mass is varied from
$M_m = 0.01$ to $10~M_{\rm Moon}$, covering the range from captured
irregular satellites to super-lunar bodies. The lower end of this
range ($0.01~M_{\rm Moon}$, comparable to the mass of Ceres) is
representative of the smaller circumplanetary debris masses produced
in SPH simulations of giant impacts, while the lunar-to-super-lunar
range brackets the debris-disk masses that typically go on to form
major satellites. The initial satellite semi-major axis is set to
$a_{m,0} = 5~R_V$ as a fiducial value (corresponding to the orbital
period of $\sim$16~hr), with additional explorations at $a_{m,0} =
3.5$--$25~R_V$. The tidal quality factor spans $Q_V = 10$--100,
appropriate for rocky bodies \citep{goldreich1966a,efroimsky2009}. For
the CTL model, $\Delta t$ is chosen to match the fiducial constant-$Q$
dissipation strength through $Q'$ at a reference tidal frequency
defined by the fiducial configuration ($P_0 = 12$~hr, $a_{m,0} =
5~R_V$). The initial orbital eccentricity is varied from $e_0 = 0$ to
$0.5$ in the eccentric calculations.

The coupled ordinary differential equations
(Equations~\ref{eq:spin_cq} and \ref{eq:orbit_cq} for the constant-$Q$
model, and Equations~\ref{eq:torque_ctl} and \ref{eq:orbit_ctl} for
the CTL model) are integrated using a fourth-order Runge-Kutta scheme
with adaptive timestep control, where each step resolves fractional
changes of 1\% in both $\Omega_V$ and $a_m$. Integration is terminated
when the satellite crosses the Roche limit, exceeds the critical
stability radius, or reaches a maximum integration time of 4.5~Gyr.
Angular momentum conservation is verified to machine precision for the
subsystem of Venus and its satellite, with the residual matching the
expected cumulative solar torque.


\subsection{Validation Against the Earth-Moon System}
\label{sec:validation}

We validate the integration framework against the Earth-Moon system,
for which observational constraints on the present-day tidal
evolution rates are available from lunar laser ranging
\citep{touma1994b}. Adopting the Earth parameters $k_{2,\oplus} =
0.299$, $Q_\oplus = 12$, $C_\oplus/(M_\oplus R_\oplus^2) = 0.3307$,
and the Moon's present semi-major axis of 60.3~$R_\oplus$, our
integration reproduces the observed lunar recession rate of 3.82~cm
yr$^{-1}$ to within 3\% (computed: 3.69~cm yr$^{-1}$) and the
observed increase in Earth's length of day of $\sim$2.3~ms
century$^{-1}$ to within 11\% (computed: 2.05~ms century$^{-1}$).
The residuals are consistent with the contributions of core-mantle
coupling and post-glacial rebound, which are not included in the
tidal model.

As a more stringent test, we integrate the Earth-Moon system from
initial conditions of $P_{\oplus,0} = 5$~hr and $a_{m,0} =
3.5~R_\oplus$ over 4.5~Gyr. With the present-day value $Q_\oplus =
12$, the integration overshoots the Moon's current distance, reaching
70.7~$R_\oplus$ with a final spin period of 56.5~hr. This is the
expected behavior: the present $Q_\oplus$ is anomalously low due to
resonant enhancement of ocean tidal dissipation by the current
continental configuration \citep{green2019}. Using the time-averaged
value $Q_\oplus \approx 34$, the integration recovers both the Moon's
present distance (60.3~$R_\oplus$) and Earth's present spin period
(24.5~hr), confirming the correct coupling between the spin and
orbital evolution equations. We stress that reproducing the Moon's
present orbit with a single constant $Q_\oplus$ is not expected in
detail. Earth's tidal response has varied substantially over its
history, and its early value is uncertain because the duration of the
post-impact molten and partially-molten state remains debated
\citep{korenaga2023a}. Dissipation in a solidifying magma ocean can
differ from the present value by orders of magnitude, limiting early
lunar recession. Our validation therefore aims only to confirm the
correct spin-orbit coupling and angular momentum methodology, not to
model Earth's detailed thermal history. Angular momentum conservation
is verified to machine precision over the full integration, with the
total change in spin-plus-orbital angular momentum matching the
expected cumulative solar torque to better than 0.1\%.

Finally, we confirm that the constant-$Q$ and CTL models produce
identical torques and migration rates when the time lag $\Delta t$ is
chosen to match $Q$ at the relevant tidal frequency, establishing the
internal consistency of the two formulations. These tests validate the
integration framework for application to the Venus-moon system, where
we explore parameter space dependence rather than fit to a specific
endpoint.


\section{Results}
\label{sec:results}


\subsection{Torque Hierarchy and Timescales}
\label{sec:timescales}

Before presenting the numerical results, it is instructive to
establish the relative magnitudes of the tidal torques. At the
fiducial configuration ($P_0 = 12$~hr, $M_m = 1~M_{\rm Moon}$, $a_{m,0} =
5~R_V$, $Q_V = 50$), the moon's tidal torque on Venus exceeds the
solar tidal torque by a factor of $\sim$$3 \times 10^6$. This stands
in stark contrast to the hot Jupiter case studied by
\citet{barnes2002b}, where the stellar torque dominates the satellite
despinning. For Venus, the moon itself is the primary agent driving
the spin evolution of the planet. The solar torque becomes relevant
only after the moon has been destroyed or has reached a
quasi-equilibrium state with Venus, at which point the solar torque
operates on a timescale of $\sim$20~Gyr for $Q_V = 50$.

A critical parameter governing the system's evolution is the
relationship between the synchronous radius and the moon's orbital
distance. For the fiducial initial semi-major axis of $a_{m,0} =
5~R_V$, the corresponding Keplerian orbital period is $P_{\rm crit} =
16.1$~hr. If Venus's initial spin period is shorter than $P_{\rm crit}$,
the synchronous radius lies inside the moon's orbit and the moon
begins in the regime of outward tidal migration, analogous to the
Earth-Moon system. If Venus's spin period exceeds $P_{\rm crit}$, the
moon starts inside the synchronous radius and immediately spirals
inward toward the Roche limit.


\subsection{Evolutionary Tracks: Constant-$Q$ Model}
\label{sec:tracks_cq}

\begin{figure*}
  \includegraphics[angle=270,width=\linewidth]{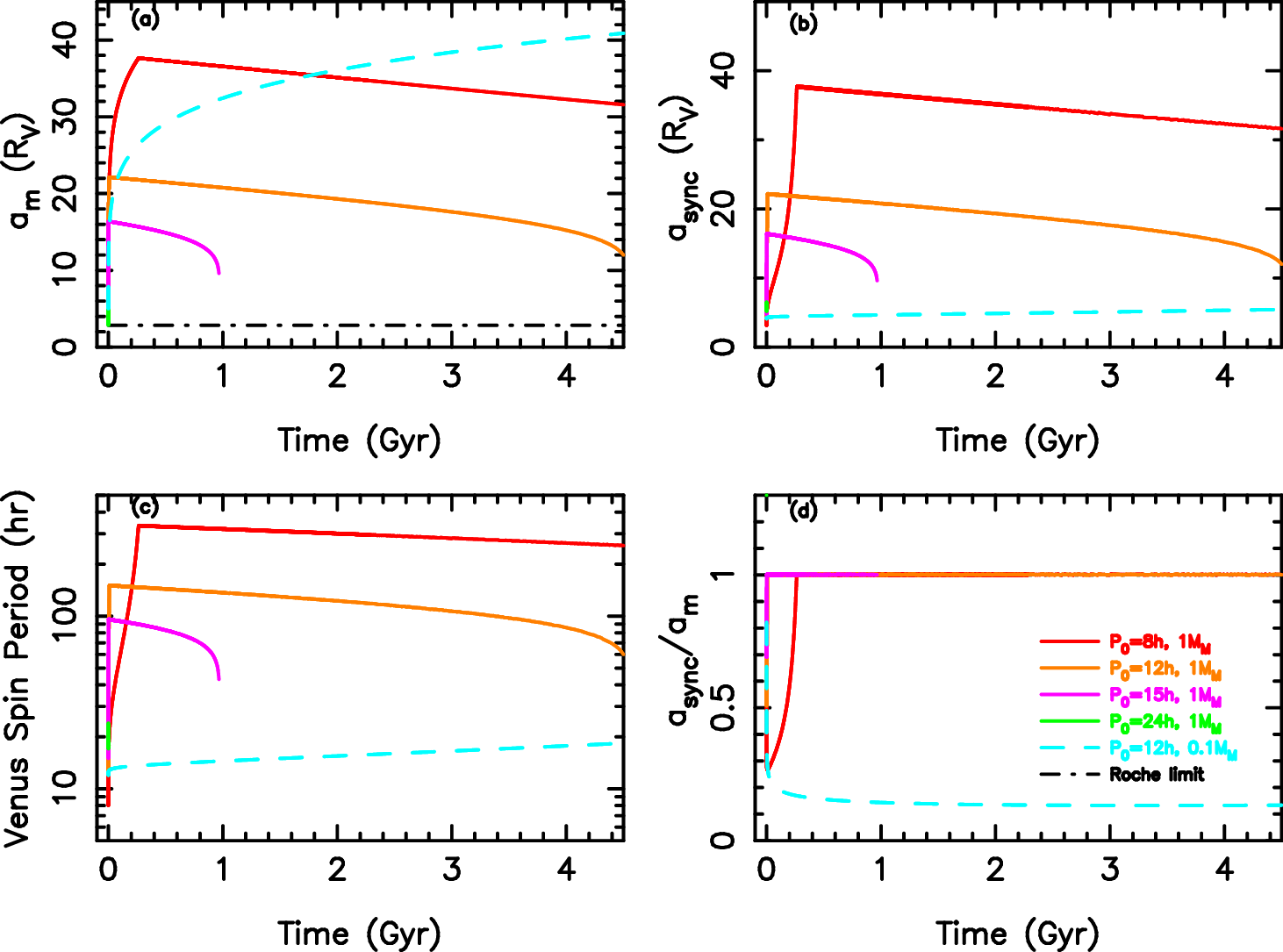}
  \caption{Tidal evolution of a hypothetical Venus moon in the
    constant-$Q$ model ($Q_V = 50$, $a_{m,0} = 5~R_V$). Panel (a)
    shows the moon's semi-major axis, panel (b) the synchronous
    radius, panel (c) the Venus spin period, and panel (d) the ratio
    of synchronous radius to orbital distance. Five cases are shown:
    $P_0 = 8$~hr (red), 12~hr (orange), and 15~hr (magenta) with
    $M_m = 1~M_{\rm Moon}$, $P_0 = 24$~hr with $M_m = 1~M_{\rm Moon}$
    (green, destroyed at the Roche limit within $\sim$1~Myr), and
    $P_0 = 12$~hr with $M_m = 0.1~M_{\rm Moon}$ (cyan dashed). The
    dash-dot line in panel (a) marks the Roche limit. For $P_0 \leq
    12$~hr, the 1~$M_{\rm Moon}$ satellite survives for 4.5~Gyr,
    while the $P_0 = 15$~hr case undergoes synchronous reversal and
    is destroyed at $\sim$0.97~Gyr.}
  \label{fig:tracks}
\end{figure*}

Figure~\ref{fig:tracks} shows representative evolutionary tracks for
the constant-$Q$ model with $Q_V = 50$ and $a_{m,0} = 5~R_V$. Two
distinct regimes are evident. For initial spin periods above $P_{\rm
crit}$ ($P_0 = 24$~hr), the moon starts inside the
synchronous radius and spirals inward to the Roche limit within
$\sim$1~Myr. The moon's inward spiral transfers orbital angular
momentum to Venus's spin, moving the synchronous radius inward, but
this feedback is insufficient to rescue the moon before it reaches the
Roche limit.

For initial spin periods below $P_{\rm crit}$ ($P_0 = 8$ and 12~hr),
the moon initially migrates outward as tidal torques transfer angular
momentum from Venus's rotation to the orbit. The same torques despin
Venus, causing the synchronous radius to expand outward. For a
1~$M_{\rm Moon}$ satellite, the outward migration is sufficiently rapid
that the moon remains ahead of the expanding synchronous radius,
reaching $\sim$25~$R_V$ before beginning a gradual inward drift as
Venus's spin continues to slow. At $Q_V = 50$, this inward drift is
too slow for the moon to reach the Roche limit within 4.5~Gyr, and
the moon survives. For a more massive moon (2~$M_{\rm Moon}$), the
stronger torque despins Venus more rapidly, causing the synchronous
radius to overtake the moon's orbit. At $P_0 = 8$~hr and $Q_V = 50$,
this reversal occurs and the moon reaches the Roche limit at
$\sim$1.7~Gyr; at $P_0 = 12$~hr the stronger initial torque drives
destruction within $\sim$33~Myr.

\begin{figure}
  \includegraphics[angle=270,width=\linewidth]{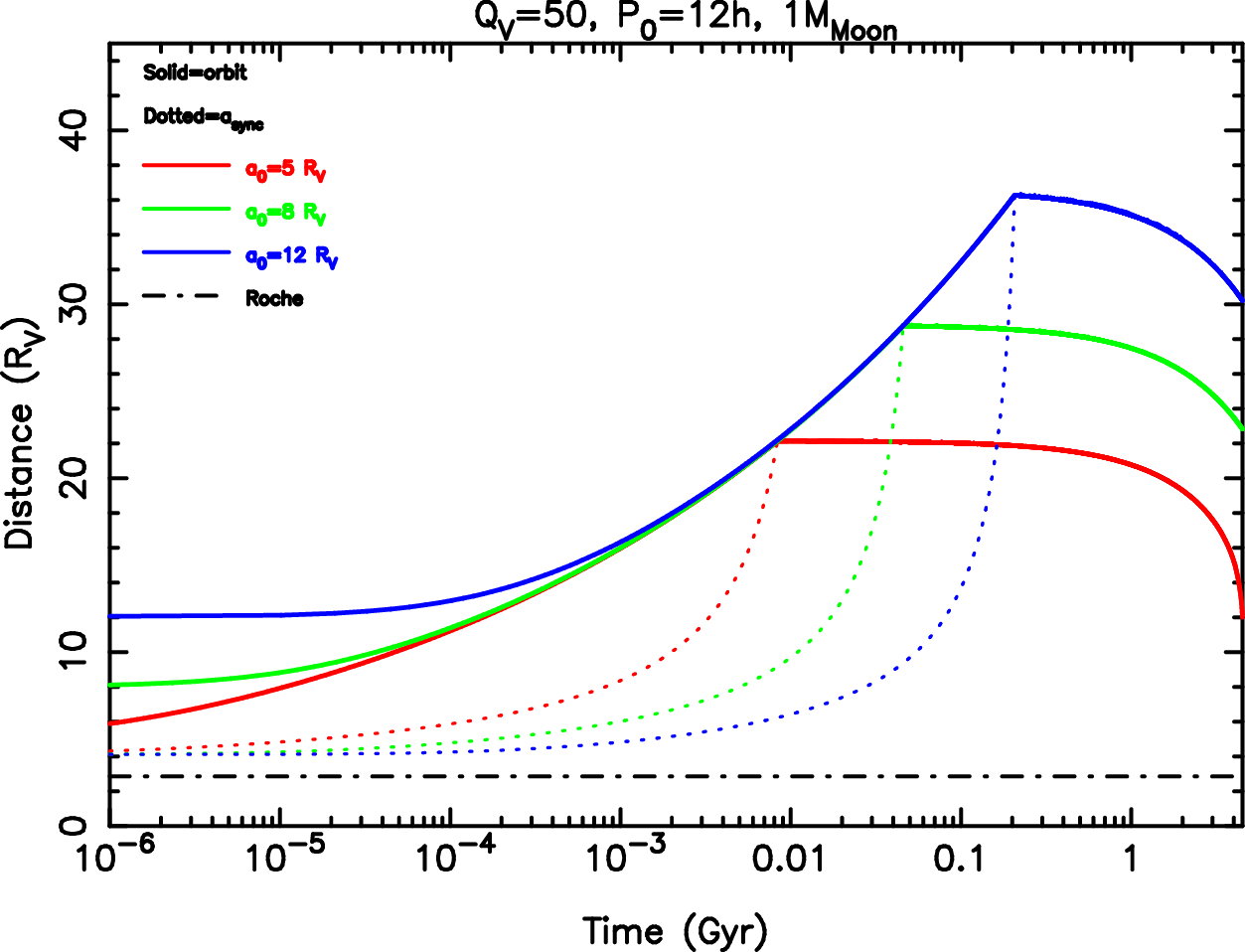}
  \caption{Dependence of the tidal evolution on the initial
    semi-major axis ($Q_V = 50$, $P_0 = 12$~hr, $M_m =
    1~M_{\rm Moon}$). Solid lines show the moon's orbital distance
    and dotted lines show the corresponding synchronous radius for
    $a_{m,0} = 5$ (red), 8 (green), and 12~$R_V$ (blue). Moons
    starting closer to Venus experience stronger tidal torques and
    migrate outward faster, but also despin Venus more rapidly. All
    three cases survive for the full 4.5~Gyr integration.}
  \label{fig:sma}
\end{figure}

The effect of satellite mass is evident from
Table~\ref{tab:lifetimes}: for a 5~$M_{\rm Moon}$ satellite, the
intense torque despins Venus almost immediately, and the moon is
destroyed within $\sim$100~Myr even at $P_0 = 5$~hr. For a less
massive moon (0.1~$M_{\rm Moon}$), the torque is too weak to despin
Venus appreciably, and the moon survives for the full 4.5~Gyr
integration, as shown by the cyan dashed curve in
Figure~\ref{fig:tracks}. Figure~\ref{fig:sma} shows the dependence on the
initial semi-major axis for a 1~$M_{\rm Moon}$ satellite at $P_0 =
12$~hr: moons starting at $a_{m,0} = 5$, 8, and 12~$R_V$ all survive
but reach different peak distances, reflecting the $a^{-6}$ scaling of
the tidal torque.

\subsection{Parameter Space: Constant-$Q$ Model}
\label{sec:param_cq}

\begin{figure*}
  \includegraphics[angle=270,width=\linewidth]{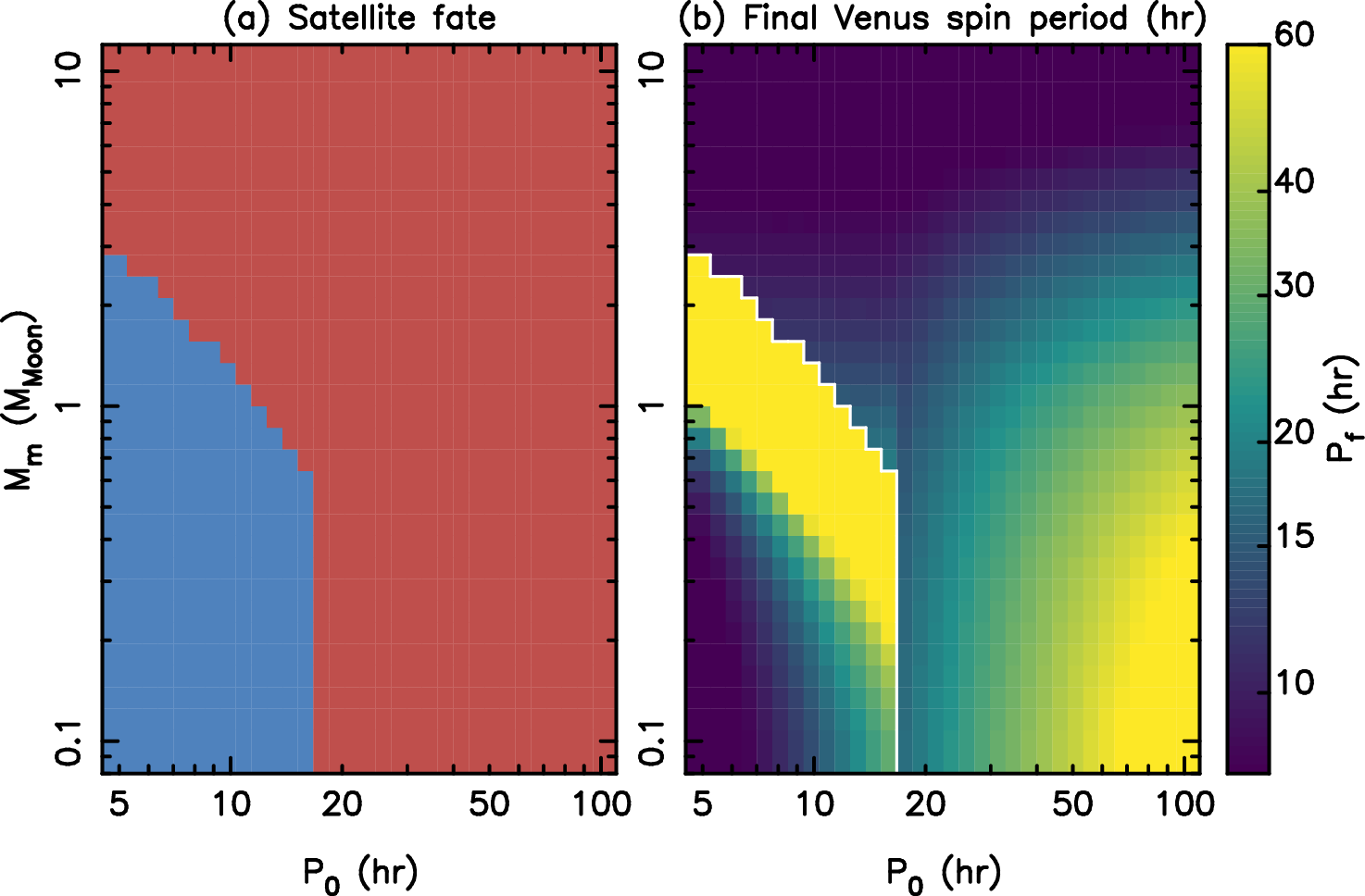}
  \caption{Two-constraint view of the constant-$Q$ model ($Q_V = 50$,
    $a_{m,0} = 5~R_V$), as a function of initial spin period and
    satellite mass. {\it Left (a):} satellite fate, with red
    indicating tidal destruction at the Roche limit and blue
    indicating survival for the full 4.5~Gyr integration. {\it Right
      (b):} Venus's final spin period (color scale), with the
    moon-loss boundary from panel (a) overdrawn as a heavy white
    line. The two panels together display the competition between the
    two constraints required to explain the present state of Venus:
    the same angular-momentum transfer that despins Venus (driving the
    final spin period to larger values, brighter colors) requires the
    moon to survive and migrate outward, whereas the parameter
    combinations that destroy the moon (left of and below the white
    boundary) leave Venus rapidly rotating (darker colors). The narrow
    region in which a moon is lost while Venus is also substantially
    despun illustrates why satisfying both constraints simultaneously
    is restrictive. The diagonal survival/destruction boundary
    reflects the competition between the moon's outward migration rate
    ($\propto M_m$) and the synchronous radius expansion rate
    ($\propto M_m^2$).}
  \label{fig:param}
\end{figure*}

Figure~\ref{fig:param} presents a two-constraint view of the ($P_0,
M_m$) parameter space for the constant-$Q$ model at the fiducial $Q_V
= 50$. Panel (a) shows the satellite fate, and panel (b) shows Venus's
final spin period with the moon-loss boundary overlaid. The figure
plainly reveals the tension between the two conditions required to
explain the present state of Venus.  Destroying the moon requires the
parameter combinations to the left of and below the
survival/destruction boundary, but these same combinations return
angular momentum to the planet and leave it rapidly rotating, whereas
substantial despinning of Venus occurs only in the regions where the
moon survives and continues to migrate outward. The results show a
sharp boundary between survival and destruction that depends on both
the satellite mass and spin period. For $Q_V = 50$, moons with $M_m
\lesssim 1~M_{\rm Moon}$ survive for the full 4.5~Gyr at spin periods
below $P_{\rm crit}$, while moons with $M_m \gtrsim 2~M_{\rm Moon}$
are destroyed through synchronous reversal on timescales of
$\sim$0.03--1.7~Gyr depending on mass and spin period. For spin
periods above $P_{\rm crit}$, all moons are destroyed within
$\lesssim$1~Myr regardless of mass. The dependence on $Q_V$ is shown
in Table~\ref{tab:lifetimes}.  Increasing $Q_V$ to 100 extends
lifetimes by a factor of $\sim$2 and shifts the survival boundary to
higher masses, while decreasing $Q_V$ to 10 compresses lifetimes and
shifts destruction to lower masses.

Table~\ref{tab:lifetimes} summarizes the moon lifetime across the
($P_0$, $M_m$) parameter space for three values of $Q_V$ in the
constant-$Q$ model. The diagonal boundary between survival and
destruction reflects the competition between the moon's outward
migration rate ($\propto M_m$) and the synchronous radius expansion
rate ($\propto M_m^2$): more massive moons despin Venus
disproportionately faster than they migrate outward, causing the
synchronous radius to overtake their orbits.

\begin{deluxetable*}{lccccccc}
\tablecaption{Moon Lifetime (Gyr) in the Constant-$Q$ Model
  ($a_{m,0} = 5~R_V$)\label{tab:lifetimes}}
\tablewidth{0pt}
\tablehead{
  \colhead{$P_0$ (hr)} &
  \multicolumn{7}{c}{$M_m$ ($M_{\rm Moon}$)} \\
  \colhead{} &
  \colhead{0.01} & \colhead{0.1} & \colhead{0.5} &
  \colhead{1.0} & \colhead{2.0} & \colhead{5.0} & \colhead{10.0}
}
\startdata
\cutinhead{$Q_V = 10$}
5  & $>$4.5 & $>$4.5 & $>$4.5 & $>$4.5 & 3.7  & 0.021 & $6 \times 10^{-4}$  \\
8  & $>$4.5 & $>$4.5 & $>$4.5 & 3.2  & 0.33 & $8 \times 10^{-4}$ & $9 \times 10^{-5}$  \\
12 & $>$4.5 & $>$4.5 & 2.5  & 0.91 & 0.007 & $6 \times 10^{-5}$ & $3 \times 10^{-5}$  \\
15 & $>$4.5 & $>$4.5 & 1.6  & 0.19 & $3 \times 10^{-4}$ & $1 \times 10^{-5}$ & $1 \times 10^{-5}$  \\
20 & R     & R     & R     & R    & R    & R    & R     \\
24 & R     & R     & R     & R    & R    & R    & R     \\
\cutinhead{$Q_V = 50$}
5  & $>$4.5 & $>$4.5 & $>$4.5 & $>$4.5 & $>$4.5 & 0.10  & 0.003  \\
8  & $>$4.5 & $>$4.5 & $>$4.5 & $>$4.5 & 1.7  & 0.004 & $5 \times 10^{-4}$  \\
12 & $>$4.5 & $>$4.5 & $>$4.5 & $>$4.5 & 0.033 & $3 \times 10^{-4}$ & $1 \times 10^{-4}$  \\
15 & $>$4.5 & $>$4.5 & $>$4.5 & 0.97 & 0.001 & $6 \times 10^{-5}$ & $7 \times 10^{-5}$  \\
20 & R     & R     & R     & R    & R    & R    & R     \\
24 & R     & R     & R     & R    & R    & R    & R     \\
\cutinhead{$Q_V = 100$}
5  & $>$4.5 & $>$4.5 & $>$4.5 & $>$4.5 & $>$4.5 & 0.21  & 0.006  \\
8  & $>$4.5 & $>$4.5 & $>$4.5 & $>$4.5 & 3.3  & 0.008 & $9 \times 10^{-4}$  \\
12 & $>$4.5 & $>$4.5 & $>$4.5 & $>$4.5 & 0.065 & $6 \times 10^{-4}$ & $3 \times 10^{-4}$  \\
15 & $>$4.5 & $>$4.5 & $>$4.5 & 1.9  & 0.003 & $1 \times 10^{-4}$ & $1 \times 10^{-4}$  \\
20 & R     & R     & R     & R    & R    & R    & R     \\
24 & R     & R     & R     & R    & R    & R    & R     \\
\enddata
\tablecomments{Entries marked ``$>$4.5'' indicate survival for the
  full 4.5~Gyr integration. Entries marked ``R'' indicate immediate
  inward spiral to the Roche limit ($P_0 > P_{\rm crit}$). Numerical
  values give the time of Roche limit crossing in Gyr. The critical
  spin period for $a_{m,0} = 5~R_V$ is $P_{\rm crit} = 16.1$~hr.}
\end{deluxetable*}

\subsection{Comparison with Earth-Moon System}
\label{sec:comparison}

\begin{figure*}
  \includegraphics[angle=270,width=\linewidth]{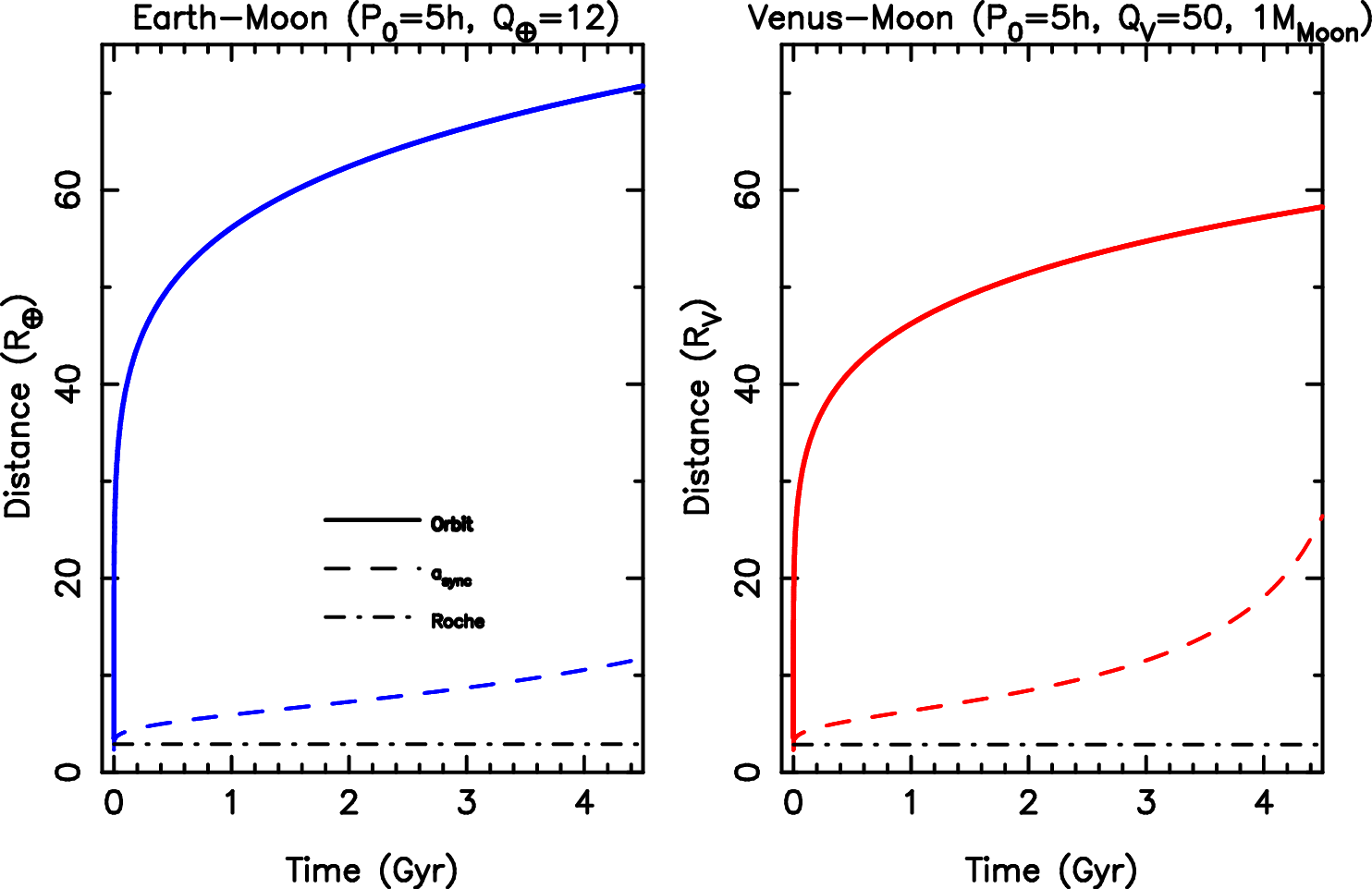}
  \caption{Comparison of the Earth-Moon system (left, blue) and an
    analogous Venus-moon system (right, red), both starting from $P_0
    = 5$~hr, $a_{m,0} = 3.5~R_p$, and $M_m = 1~M_{\rm Moon}$. Solid
    lines show the moon's orbital distance, dashed lines show the
    synchronous radius, and dash-dot lines show the Roche limit. For
    Earth ($Q_\oplus = 12$), the Moon migrates steadily outward to
    60~$R_\oplus$. For Venus ($Q_V = 50$), the moon also migrates
    outward but the synchronous radius grows faster, narrowing the
    margin between orbit and synchronization.}
  \label{fig:compare}
\end{figure*}

Figure~\ref{fig:compare} directly compares the tidal evolution of the
Earth-Moon system with the analogous Venus-moon system, both starting
from a 5-hour spin period and 3.5~$R_p$ initial semi-major axis.
We adopt $Q_\oplus = 12$, as constrained by lunar laser ranging
measurements of the present-day recession rate \citep{touma1994b},
and $Q_V = 50$, a canonical value for rocky terrestrial bodies
\citep{goldreich1966a} that is appropriate given the absence of any direct
observational constraint on Venus's tidal dissipation. The use of
planet-specific $Q$ values is deliberate because the comparison is intended
to reflect the actual physical environments of each system rather
than isolate a single variable. For
Earth, the Moon migrates steadily outward for the full 4.5~Gyr,
reaching $\sim$60~$R_\oplus$, because the synchronous radius remains
well inside the Moon's orbit throughout. This is a consequence of
Earth's retention of substantial spin angular momentum, mediated in
part by the relatively weak solar tidal torque at 1~AU.

For Venus, the moon also migrates outward, but the stronger tidal
torque (due to Venus's smaller Hill sphere requiring the moon to orbit
closer) combined with the stronger solar tidal torque (due to Venus's
proximity to the Sun at 0.72~AU) drives more rapid despinning. The
synchronous radius expands faster for Venus than for Earth, gradually
closing the gap with the moon's orbit. While the 1~$M_{\rm Moon}$
satellite survives at $P_0 = 5$~hr, the margin of safety is much
smaller than for Earth: the synchronous radius reaches $\sim$50\% of
the moon's distance by 4.5~Gyr, compared with $\sim$15\% for Earth.
This fundamental asymmetry between Earth and Venus arises from the
$a^{-6}$ dependence of the tidal torque: a moon forced to orbit closer
exerts a far stronger torque, creating a positive feedback loop between
despinning and synchronous radius expansion that, for slightly slower
initial spin or more massive moons, leads to orbital decay and
destruction. Thus, while the specific initial conditions of the Moon-forming
impact would have permitted satellite survival at Venus, the
parameter space analysis of Section~\ref{sec:param_cq} demonstrates
that this outcome occupies a narrow region; for the majority of
plausible post-impact spin periods and satellite masses, the Venus
system crosses into the destruction regime.

\subsection{Constant-$Q$ versus CTL Model}
\label{sec:cq_vs_ctl}

\begin{figure*}
  \includegraphics[angle=270,width=\linewidth]{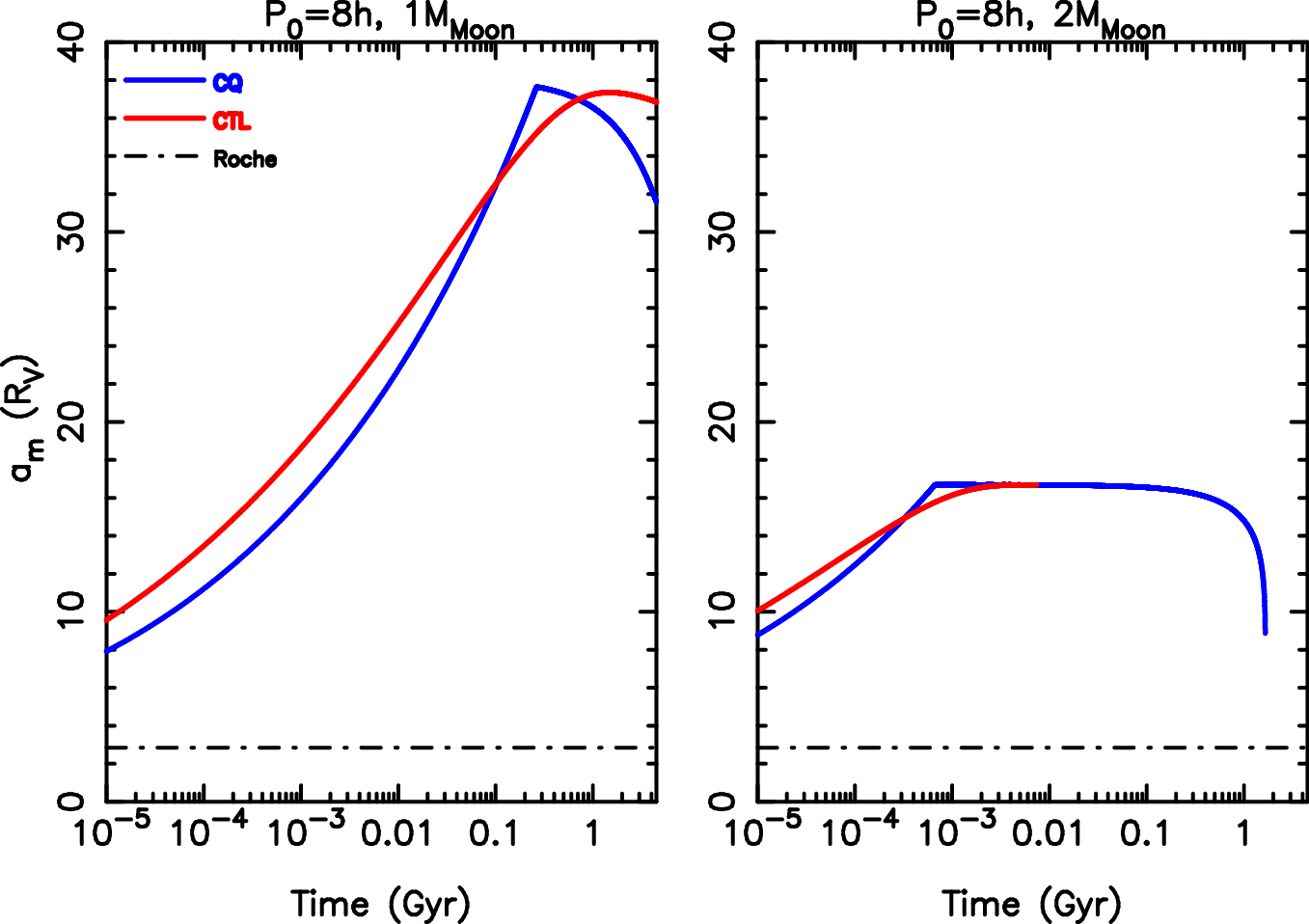}
  \caption{Comparison of the constant-$Q$ (blue) and CTL (red) tidal
    models for $P_0 = 8$~hr and $a_{m,0} = 5~R_V$. Left panel: $M_m =
    1~M_{\rm Moon}$, where both models agree and the moon survives.
    Right panel: $M_m = 2~M_{\rm Moon}$, where the models diverge in
    that the constant-$Q$ model predicts Roche destruction at
    $\sim$1.7~Gyr, while the CTL model predicts survival through
    quasi-synchronous equilibrium. The dash-dot line marks the Roche
    limit.}
  \label{fig:cqctl}
\end{figure*}

Figure~\ref{fig:cqctl} presents a direct comparison of the
constant-$Q$ and CTL models for two cases at $P_0 = 8$~hr that
isolate the effect of satellite mass. The two
models agree for a 1~$M_{\rm Moon}$ satellite,
where both predict survival through quasi-synchronous equilibrium.

The models diverge for massive moons ($M_m \gtrsim 2~M_{\rm Moon}$) at
spin periods below $P_{\rm crit}$. In the constant-$Q$ model, the
strong torque from a massive moon despins Venus rapidly, the
synchronous radius overtakes the moon's orbit, and the torque reversal
drives inward spiral to Roche destruction. In the CTL model, the torque
weakens as $(\Omega_V - n_m) \rightarrow 0$, and the system approaches
a tidal fixed point at the synchronous configuration. The moon
asymptotically approaches a quasi-synchronous state analogous to the
Pluto-Charon double tidal lock \citep{peale1999a}, extending its
lifetime beyond the age of the Solar System.

Quantitatively, for $P_0 = 8$~hr and $M_m = 2~M_{\rm Moon}$, the
constant-$Q$ model predicts Roche destruction at 1.7~Gyr while the CTL
model predicts survival. For $P_0 = 12$~hr and the same mass, the
constant-$Q$ model gives destruction at 33~Myr while the CTL model
again predicts survival. For $M_m = 5~M_{\rm Moon}$, the constant-$Q$
model drives destruction within $\sim$100~Myr at all spin periods,
while the CTL model still finds quasi-synchronous equilibrium for $P_0
\leq 15$~hr. These differences represent qualitatively distinct
outcomes that depend on the assumed tidal rheology.

\subsection{Effect of Initial Eccentricity}
\label{sec:ecc}

\begin{figure*}
  \includegraphics[angle=270,width=\linewidth]{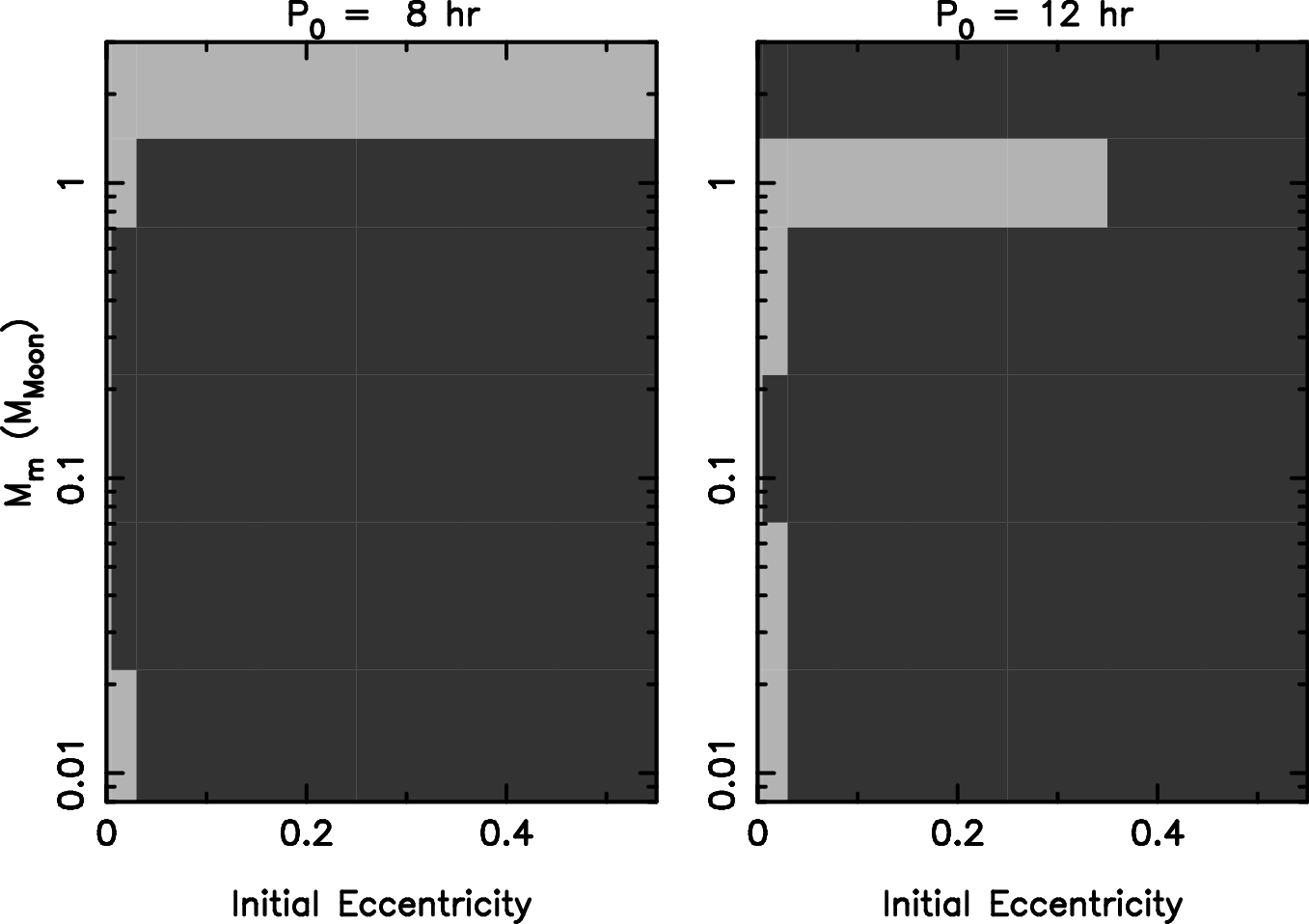}
  \caption{Moon survival in the CTL eccentric model as a function of
    initial eccentricity and satellite mass ($a_{m,0} = 5~R_V$,
    $Q_{\rm eff} \approx 50$). Gray regions indicate survival for
    4.5~Gyr; dark regions indicate destruction either through Roche
    limit crossing (inward spiral) or Hill sphere escape
    (eccentricity-driven outward migration). Left panel: $P_0 = 8$~hr,
    where $\Omega_V/n_m = 2.0 > 18/11$ places the system initially in
    the eccentricity-pumping regime. Massive moons despin Venus below
    the pumping threshold before eccentricity grows, while low-mass
    moons remain in the pumping regime and are destabilized. Right
    panel: $P_0 = 12$~hr, where $\Omega_V/n_m = 1.3 < 18/11$ and
    eccentricity is damped, permitting survival near $M_m \approx
    1~M_{\rm Moon}$ at $e_0 \lesssim 0.3$; destruction at higher $e_0$
    is through enhanced tidal dissipation driving Roche limit
    crossing. For $P_0 = 24$~hr (not shown), the moon starts inside
    the synchronous radius and reaches the Roche limit regardless of
    eccentricity.}
  \label{fig:ecc_param}
\end{figure*}

We explore the effect of initial orbital eccentricity using the full
Hut tidal equations \citep{hut1981c,leconte2010a}, which are exact in
the viscous approximation for arbitrary eccentricity.
Figure~\ref{fig:ecc_param} maps the moon survival as a function of
initial eccentricity and moon mass for $P_0 = 8$ and 12~hr.

The eccentricity evolution is governed by a critical frequency ratio:
in the small-eccentricity limit, the Hut $de/dt$ equation changes sign
at $\Omega_V/n_m = 18/11 \approx 1.636$ \citep{hut1981c}. At finite
eccentricity, the threshold shifts as the eccentricity functions
$\Omega_e(e)$ and $N_e(e)$ diverge from unity; the full $e$-dependent
condition is used in our numerical integrations. When Venus spins
faster than this threshold, tidal torques pump eccentricity rather
than damping it. For the fiducial initial semi-major axis of $a_{m,0}
= 5~R_V$, the small-$e$ threshold corresponds to $P_0 \approx 10$~hr:
spin periods shorter than $\sim$10~hr place the system in the
eccentricity-pumping regime.

For $P_0 = 8$~hr ($\Omega_V/n_m = 2.0$), the system begins in the
eccentricity-pumping regime for all satellite masses. However, the
outcome depends critically on how quickly Venus despins below the
$18/11$ threshold. Massive moons ($M_m \gtrsim 2~M_{\rm Moon}$) exert
strong enough torques to despin Venus to quasi-synchronous lock before
eccentricity can grow appreciably; once $\Omega_V/n_m$ drops below
$18/11$, eccentricity damps and the moon survives. Low-mass moons
($M_m \lesssim 0.5~M_{\rm Moon}$), by contrast, leave Venus spinning
super-synchronously for Gyr timescales. The growing eccentricity
enhances the $(1-e^2)^{-15/2}$ dissipation factors, which accelerate
outward migration and further eccentricity growth in a positive
feedback loop. The moon is driven toward the Hill sphere stability
boundary and lost from the system, even for initial eccentricities as
small as $e_0 = 0.01$.

For $P_0 = 12$~hr ($\Omega_V/n_m = 1.3$), eccentricity is damped on
timescales shorter than the spin evolution timescale, and the system
approaches the same tidal equilibrium as the circular case. Moderate
initial eccentricities ($e_0 \leq 0.3$) do not qualitatively alter the
survival boundary near $M_m \approx 1~M_{\rm Moon}$, though lower-mass
moons are destabilized at smaller $e_0$. At $e_0 \gtrsim 0.5$, the
enhanced dissipation from the $(1-e^2)^{-15/2}$ factors overwhelms the
synchronous attractor even in the damping regime, driving the moon to
destruction.

For $P_0 = 24$~hr (not shown in Figure~\ref{fig:ecc_param}), the
moon starts inside the synchronous radius and eccentricity has no
effect on the outcome: the grid calculations confirm that all cases
result in rapid Roche destruction regardless of $e_0$.

The eccentricity pumping criterion adds a mass-dependent constraint
on the survival of a hypothetical Venus moon. For fast initial spin
($P_0 \lesssim 10$~hr), lunar-mass and sub-lunar satellites are
vulnerable to eccentricity-driven destabilization even at modest
$e_0$, while more massive moons can despin Venus out of the pumping
regime before eccentricity grows.

\section{Discussion}
\label{sec:disc}


\subsection{The Rheological Discriminant}
\label{sec:rheology}

The main result of this work is that the survival of a hypothetical
Venus moon depends critically on the rheological response of Venus's
interior at low tidal forcing frequencies. The constant-$Q$ model,
which assumes frequency-independent dissipation, predicts moon
destruction for $M_m \gtrsim 2~M_{\rm Moon}$ within
$\sim$0.03--1.7~Gyr through synchronous reversal. The CTL model, which
assumes dissipation proportional to tidal frequency, predicts survival
for the same masses when the initial spin period is shorter than
$P_{\rm crit}$. For lower-mass moons ($M_m \lesssim 1~M_{\rm Moon}$),
both models agree that survival is possible at fast spin periods.  Of
the two models, the CTL framework is generally considered more
physically motivated near the synchronous configuration, because the
constant-$Q$ torque reverses discontinuously at $\Omega_V = n_m$ while
the CTL torque passes smoothly through zero, consistent with the
observed tidal locking of systems such as Pluto-Charon
\citep{peale1999a}. Indeed, the present-day retrograde rotation of
Venus has been modeled as an equilibrium between gravitational body
tides and thermal atmospheric tides \citep{correia2001,correia2003a},
underscoring the role of frequency-dependent dissipation in governing
Venus's long-term spin evolution. We note that the constant-$Q$
lifetimes in Table~\ref{tab:lifetimes} are sensitive to the numerical
treatment of the torque sign reversal at $\Omega_V = n_m$, because the
discontinuous sign flip produces oscillations whose net effect depends
on the integration resolution. The CTL model, which transitions
smoothly through synchronization, provides a more robust estimate of
lifetimes near the survival boundary. This behavior has a physical
basis: the synchronous state is formally a spin-orbit resonance
\citep{goldreich1966f}, and the coupled evolution of $\Omega_V$ and
$a_m$ near $\Omega_V = n_m$ produces a restoring feedback analogous to
resonant libration. The tidal torque that moves the satellite also
adjusts Venus's spin rate, driving the synchronous radius toward the
satellite's orbit and resisting departure from synchronization. The
CTL model captures this libration naturally through its smooth,
frequency-dependent torque, while the constant-$Q$ sign discontinuity
replaces the smooth transit with numerical oscillations whose
resolution depends on the integration timestep.

Neither the constant-$Q$ nor CTL model are fully realistic for a rocky
body. Laboratory measurements and geophysical constraints suggest that
rocky planet interiors exhibit a complex rheological response
intermediate between these two limits, potentially following an
Andrade or Sundberg-Cooper model \citep{efroimsky2012a,renaud2018}. At
low forcing frequencies, the Andrade model predicts weaker dissipation
than the constant-$Q$ model but stronger than the CTL model,
suggesting that the true evolution may lie between our two computed
extremes.  Future work coupling frequency-dependent tidal models to
the Venus-moon system would help narrow this range.


\subsection{Constraints from Giant Impact Simulations}
\label{sec:spin}

The survival of a lunar-mass satellite in our calculations requires
Venus's post-impact spin period to be $\lesssim$12~hr
(Table~\ref{tab:lifetimes}). This places a strong constraint on the
formation scenario: if the moon-forming impact left Venus spinning
more slowly, a lunar-mass moon would be destroyed within
$\sim$1~Gyr. The energetics of such a scenario would be dramatic:
\citet{nimmo2024c} demonstrated that tidal forces during the Moon's
early outward migration from Earth were sufficient to remelt the lunar
surface, and the tidal heating during an inward spiral toward the
Roche limit would be at least as intense.

Recent SPH simulations of giant impacts on Venus provide direct
constraints on the expected post-impact spin state.
\citet{bussmann2025} explored a range of impact scenarios including
head-on collisions on a non-rotating Venus and oblique impacts by
Mars-sized bodies, and found that scenarios consistent with Venus's
present-day rotation typically produce post-impact spin periods of
$\gtrsim$12~hr. An important aspect is that these same impact
geometries produce debris disks that reside within the synchronous
orbit, preventing moon formation entirely.

Several caveats apply to the application of these simulations to our
results. \citet{bussmann2025} did not systematically vary the
pre-impact spin state, adopted a fully differentiated initial Venus,
and were principally concerned with matching Venus's present spin
directly through the impact rather than through subsequent moon-driven
migration. Moreover, many of the smaller impactors in their suite may
not correspond to a genuine ``last giant impact,'' since a more
massive impactor would plausibly have occurred earlier in the
accretion history. Only the subset of their simulations that produce a
rapidly-rotating, debris-disk-bearing post-impact Venus is directly
comparable to the initial conditions modeled here. The remainder
represent a distinct formation pathway in which the moon-survival
question does not arise. We therefore treat the \citet{bussmann2025}
spin-period estimates as indicative rather than definitive, and note
that a dedicated study coupling the full range of post-impact states
to tidal evolution would be required to draw firm conclusions.

This result establishes a tension for the moon survival scenario.
Impact geometries that produce fast post-impact spin ($P_0 < 12$~hr)
would place a lunar-mass moon in the survival zone of our tidal
evolution calculations, but such rapid rotation may be inconsistent
with Venus's subsequent spin evolution to its present retrograde
state. Conversely, impact geometries that produce slower post-impact
spin ($P_0 > 15$~hr) are more compatible with Venus's present
rotation but place any moon firmly in the destruction zone. The narrow
window at $P_0 \approx 12$--$15$~hr, where a lunar-mass moon is
marginally destroyed on Gyr timescales, may represent the most
physically relevant regime for Venus's actual history.

The combination of our tidal analysis with these impact constraints
suggests that Venus's lack of a moon can be attributed to one of two
mechanisms: either the moon-forming impact geometry prevented the
formation of a stable circumplanetary disk in the first place
\citep{bussmann2025}, or a moon did form but was tidally destroyed on
a timescale set by the post-impact spin period. In either case, the
absence of a Venusian satellite is a natural consequence of the
initial conditions rather than requiring a separate catastrophic event.


\subsection{Implications for Early Venus Habitability}
\label{sec:hab}

The presence of a large satellite during Venus's first $\sim$1~Gyr
would have several consequences for the planet's early habitability
\citep{kane2014e,kane2024b}. A moon of
$\sim$1~$M_{\rm Moon}$ would stabilize Venus's obliquity against
chaotic variations
\citep{laskar1993b,lissauer2012a}, potentially maintaining favorable
insolation patterns for surface habitability. Tidal interactions
between Venus and a close-in satellite would drive ocean tides of
substantial amplitude \citep{green2019}, which in turn
influence rotational evolution and heat transport in a putative early
ocean. The eventual destruction of the moon at the Roche limit would
deliver a massive energy pulse to the Venusian surface and atmosphere,
with potential implications for the loss of surface water and the
onset of a runaway greenhouse state.

In the CTL model, where the moon can survive for Gyr timescales, the
window for these habitability benefits extends through the period when
Venus may have retained surface water
\citep{way2020}. The eventual loss of this
stabilizing influence, whether through tidal destruction or secular
perturbation, could represent one of the triggers for Venus's
transition from a potentially habitable state to its present
inhospitable condition.

We note that our tidal evolution calculations include only
gravitational body tides from both the satellite and the Sun, and do
not account for atmospheric thermal tides. For Venus, atmospheric
tides driven by solar heating are thought to play a significant role
in the planet's spin evolution
\citep{correia2001,correia2003a,leconte2015a,margot2021c,kane2022b}, providing an
additional despinning torque that would accelerate the expansion of
the synchronous radius. Including this effect would hasten the
onset of inward migration and reduce the satellite lifetime relative
to our computed values, further narrowing the survival region in
parameter space. The omission of atmospheric tides therefore results in
conservative values, since the true survival prospects for a Venusian moon are
likely even more limited than the results presented here suggest.


\subsection{Implications for Venus Zone Exoplanets}
\label{sec:exo}

We note that the models in this work are calibrated specifically to
Venus's mass, radius, and orbital distance, and the quantitative
results should not be read as a general survey of Venus-analog
systems. The purpose of this subsection is the more limited one of
indicating the direction in which the Venus results generalize; a
comprehensive treatment, varying stellar mass, orbital separation, and
the insolation-dependent tidal response, is left to future work.  With
that caveat in mind, there are qualitative implications for
terrestrial exoplanets in the Venus Zone
\citep{kane2014e,kane2019d,ostberg2023a}. Slowly rotating terrestrial
planets orbiting interior to the habitable zone are expected to have
spin periods longer than $P_{\rm crit}$ for any plausible satellite,
placing them in the regime of immediate inward migration and rapid
moon destruction \citep{kane2022b}. This suggests that Venus-analog
exoplanets should generically lack large satellites \citep{kane2017c},
with consequences for obliquity stability and habitability
assessments.

For Venus-analog planets orbiting M~dwarfs, the situation is even more
extreme. The reduced Hill sphere radius (due to the closer orbital
distance required for equivalent insolation) forces satellites into
tighter orbits, amplifying the $a^{-6}$ tidal torque and accelerating
their destruction \citep{barnes2002b}. Quantitatively, for a moon at a
fixed fraction $f$ of the planetary Hill sphere, the critical spin
period scales as
\begin{equation}
  P_{\rm crit} \propto f^{3/2} \, a_{\rm orb}^{3/2} \, M_\star^{-1/2}
  \label{eq:pcrit_scaling}
\end{equation}
where $a_{\rm orb}$ is the planet's orbital semi-major axis and
$M_\star$ is the stellar mass. For a Venus-analog planet at 0.1~AU
around a 0.3~$M_\odot$ M~dwarf, $P_{\rm crit}$ is reduced by a
factor of $\sim$10 relative to Venus, rendering moon survival
effectively impossible for any plausible post-impact spin period. The
habitability of such worlds cannot rely on the obliquity-stabilizing
influence of a large moon.


\subsection{Connection to DAVINCI}
\label{sec:davinci}

If a Venusian moon was destroyed at the Roche limit, the resulting
debris would have been deposited onto Venus's surface and into its
atmosphere. The NASA DAVINCI mission \citep{garvin2022}, which will
perform in situ measurements of Venus's atmospheric composition during
descent, could potentially constrain this scenario through
measurements of noble gas abundances and isotopic ratios. A late
delivery of volatile-poor silicate material from a disrupted moon
would have a distinct signature compared with the indigenous volatile
inventory, although disentangling this signal from the effects of
subsequent volcanic outgassing and atmospheric escape remains a
substantial challenge. A useful next step would be a mass-balance
calculation of lunar-composition material delivered to Venus's
atmosphere and surface, which would help quantify the detectability of
such a signature against the background volatile inventory. The
connection between the tidal destruction timescales derived here and
the DAVINCI measurement objectives provides motivation for future
modeling of the atmospheric chemical consequences of a Roche limit
disruption event.

The observed age of Venus's surface provides a complementary timing
constraint. Venus's surface is globally young, with a crater-retention
age of several hundred Myr, implying substantial resurfacing well
after the epoch of the last giant impact. Because the reaccretion of a
tidally disrupted moon would deposit a large mass of silicate debris
onto the surface, such an event could not have occurred within the
most recent resurfacing interval without leaving an observable
signature. Any moon-loss event must therefore predate the surface age,
placing it early in Venus's history, consistent with the sub-Gyr tidal
destruction timescales derived here for the physically favored
parameter range. This reinforces the picture in which moon loss, if it
occurred, was an early consequence of the post-impact tidal evolution
rather than a recent event.


\section{Conclusions}
\label{sec:con}

The tidal evolution of a hypothetical Venus-moon system is governed by
a competition between the moon's outward migration rate, which scales
linearly with satellite mass, and the rate at which Venus's despinning
expands the synchronous radius, which scales as the square of the
satellite mass. This asymmetry creates a diagonal boundary in the
($P_0$, $M_m$) parameter space that separates survival from
destruction. For a lunar-mass satellite formed around a rapidly
spinning Venus ($P_0 \lesssim 12$~hr), the moon migrates outward
faster than the synchronous radius expands, and the system reaches a
quasi-synchronous equilibrium that persists for the age of the Solar
System. This is the same fundamental physics that governs the
Earth-Moon system, and is robust across both tidal models and the
explored range of $Q_V$. In the constant-$Q$ model, more massive
satellites ($M_m \gtrsim 2~M_{\rm Moon}$), or satellites around more
slowly rotating Venus ($P_0 \gtrsim 15$~hr), undergo synchronous
reversal and spiral inward to the Roche limit on timescales of
$\sim$0.03--1.7~Gyr. In the CTL model, however, massive moons at fast
initial spin can instead approach quasi-synchronous equilibrium,
emphasizing that this boundary is rheology-dependent. For spin periods
exceeding the critical value $P_{\rm crit}$, the moon starts inside
the synchronous radius and is destroyed within $\lesssim$1~Myr
regardless of mass or tidal model. The moon's tidal torque on Venus
exceeds the solar tidal torque by a factor of $\sim$$3 \times 10^6$,
making the moon itself the dominant agent of Venus's despinning; a
significant contrast with the hot Jupiter satellite problem where
stellar tides dominate. The survival window is further shaped by the
eccentricity pumping criterion: for $P_0 \lesssim 10$~hr, the
spin-to-orbit frequency ratio exceeds $18/11$ and tidal torques
initially amplify orbital eccentricity. Low-mass satellites are
destroyed through this mechanism even at $e_0$ as small as 0.01,
while massive moons can despin Venus below the pumping threshold
before eccentricity grows. For a lunar-mass satellite, survival
requires $P_0 \approx 10$--$12$~hr with low initial eccentricity.

The combination of these tidal constraints with recent giant impact
simulations paints a self-consistent picture for Venus's lack of a
moon. Impact geometries that leave Venus spinning rapidly ($P_0 <
12$~hr) would place a lunar-mass moon in the survival zone of our
calculations, but these same geometries tend to produce debris disks
inside the synchronous orbit that reaccrete without forming a moon
\citep{bussmann2025}. Conversely, impact geometries that produce
debris outside the synchronous orbit (and thus could form a moon)
typically leave Venus spinning more slowly ($P_0 \gtrsim 12$~hr),
placing the resulting satellite at the boundary between survival and
destruction. For last-impact conditions in this regime, the absence of
a Venusian satellite can be understood as a natural consequence of
tidal evolution alone. A subsequent catastrophic stripping event,
while capable of removing a moon, is not required.  This convergence
of impact and tidal constraints provides motivation for further
investigation into the atmospheric chemical consequences of a Roche
limit disruption event, including potential signatures detectable by
the DAVINCI mission \citep{garvin2022}.  The application of these
results to Venus Zone exoplanets
\citep{kane2014e,kane2019d,ostberg2023a} suggests that slowly rotating
terrestrial planets in the inner regions of planetary systems are
unlikely to retain large satellites, with important implications for
their obliquity stability and long-term habitability.


\section*{Acknowledgements}

The authors would like to thank the referee, Matthew Clement, whose
feedback helped to improve the manuscript. The results reported herein
benefited from collaborations and/or information exchange within
NASA's Nexus for Exoplanet System Science (NExSS) research
coordination network sponsored by NASA's Science Mission Directorate.



\begin{thebibliography}{}
\expandafter\ifx\csname natexlab\endcsname\relax\def\natexlab#1{#1}\fi
\providecommand{\url}[1]{\href{#1}{#1}}
\providecommand{\dodoi}[1]{doi:~\href{http://doi.org/#1}{\nolinkurl{#1}}}
\providecommand{\doeprint}[1]{\href{http://ascl.net/#1}{\nolinkurl{http://ascl.net/#1}}}
\providecommand{\doarXiv}[1]{\href{https://arxiv.org/abs/#1}{\nolinkurl{https://arxiv.org/abs/#1}}}

\bibitem[{{Barnes} \& {O'Brien}(2002)}]{barnes2002b}
{Barnes}, J.~W., \& {O'Brien}, D.~P. 2002, \apj, 575, 1087,
  \dodoi{10.1086/341477}

\bibitem[{{Bills}(1992)}]{bills1992a}
{Bills}, B.~G. 1992, \grl, 19, 1025, \dodoi{10.1029/92GL01067}

\bibitem[{{Bolmont} {et~al.}(2015){Bolmont}, {Raymond}, {Leconte}, {Hersant},
  \& {Correia}}]{bolmont2015}
{Bolmont}, E., {Raymond}, S.~N., {Leconte}, J., {Hersant}, F., \& {Correia},
  A.~C.~M. 2015, \aap, 583, A116, \dodoi{10.1051/0004-6361/201525909}

\bibitem[{{Bro{\v{z}}} {et~al.}(2021){Bro{\v{z}}}, {Chrenko}, {Nesvorn{\'y}},
  \& {Dauphas}}]{broz2021c}
{Bro{\v{z}}}, M., {Chrenko}, O., {Nesvorn{\'y}}, D., \& {Dauphas}, N. 2021,
  Nature Astronomy, 5, 898, \dodoi{10.1038/s41550-021-01383-3}

\bibitem[{{Burns}(1973)}]{burns1973b}
{Burns}, J.~A. 1973, Nature Physical Science, 242, 23,
  \dodoi{10.1038/physci242023a0}

\bibitem[{{Bussmann} {et~al.}(2025){Bussmann}, {Reinhardt}, {Gillmann},
  {Meier}, {Stadel}, {Tackley}, \& {Helled}}]{bussmann2025}
{Bussmann}, M., {Reinhardt}, C., {Gillmann}, C., {et~al.} 2025, \aap, 702,
  A106, \dodoi{10.1051/0004-6361/202555802}

\bibitem[{{Canup} {et~al.}(2001){Canup}, {Ward}, \& {Cameron}}]{canup2001a}
{Canup}, R.~M., {Ward}, W.~R., \& {Cameron}, A.~G.~W. 2001, \icarus, 150, 288,
  \dodoi{10.1006/icar.2000.6581}

\bibitem[{{Correia} \& {Laskar}(2001)}]{correia2001}
{Correia}, A. C.~M., \& {Laskar}, J. 2001, \nat, 411, 767,
  \dodoi{10.1038/35081000}

\bibitem[{{Correia} {et~al.}(2003){Correia}, {Laskar}, \& {de
  Surgy}}]{correia2003a}
{Correia}, A. C.~M., {Laskar}, J., \& {de Surgy}, O.~N. 2003, \icarus, 163, 1,
  \dodoi{10.1016/S0019-1035(03)00042-3}

\bibitem[{{{\'C}uk} {et~al.}(2016){{\'C}uk}, {Hamilton}, {Lock}, \&
  {Stewart}}]{cuk2016b}
{{\'C}uk}, M., {Hamilton}, D.~P., {Lock}, S.~J., \& {Stewart}, S.~T. 2016,
  \nat, 539, 402, \dodoi{10.1038/nature19846}

\bibitem[{{{\'C}uk} \& {Stewart}(2012)}]{cuk2012c}
{{\'C}uk}, M., \& {Stewart}, S.~T. 2012, Science, 338, 1047,
  \dodoi{10.1126/science.1225542}

\bibitem[{{Domingos} {et~al.}(2006){Domingos}, {Winter}, \&
  {Yokoyama}}]{domingos2006}
{Domingos}, R.~C., {Winter}, O.~C., \& {Yokoyama}, T. 2006, \mnras, 373, 1227,
  \dodoi{10.1111/j.1365-2966.2006.11104.x}

\bibitem[{{Efroimsky}(2012)}]{efroimsky2012a}
{Efroimsky}, M. 2012, \apj, 746, 150, \dodoi{10.1088/0004-637X/746/2/150}

\bibitem[{{Efroimsky} \& {Williams}(2009)}]{efroimsky2009}
{Efroimsky}, M., \& {Williams}, J.~G. 2009, Celestial Mechanics and Dynamical
  Astronomy, 104, 257, \dodoi{10.1007/s10569-009-9204-7}

\bibitem[{{Evans} {et~al.}(2015){Evans}, {Peplowski}, {McCubbin}, {McCoy},
  {Nittler}, {Zolotov}, {Ebel}, {Lawrence}, {Starr}, {Weider}, \&
  {Solomon}}]{evans2015b}
{Evans}, L.~G., {Peplowski}, P.~N., {McCubbin}, F.~M., {et~al.} 2015, \icarus,
  257, 417, \dodoi{10.1016/j.icarus.2015.04.039}

\bibitem[{{Farhat} {et~al.}(2022){Farhat}, {Auclair-Desrotour}, {Bou{\'e}}, \&
  {Laskar}}]{farhat2022b}
{Farhat}, M., {Auclair-Desrotour}, P., {Bou{\'e}}, G., \& {Laskar}, J. 2022,
  \aap, 665, L1, \dodoi{10.1051/0004-6361/202243445}

\bibitem[{{Garvin} {et~al.}(2022){Garvin}, {Getty}, {Arney}, {Johnson},
  {Kohler}, {Schwer}, {Sekerak}, {Bartels}, {Saylor}, {Elliott}, {Goodloe},
  {Garrison}, {Cottini}, {Izenberg}, {Lorenz}, {Malespin}, {Ravine}, {Webster},
  {Atkinson}, {Aslam}, {Atreya}, {Bos}, {Brinckerhoff}, {Campbell}, {Crisp},
  {Filiberto}, {Forget}, {Gilmore}, {Gorius}, {Grinspoon}, {Hofmann}, {Kane},
  {Kiefer}, {Lebonnois}, {Mahaffy}, {Pavlov}, {Trainer}, {Zahnle}, \&
  {Zolotov}}]{garvin2022}
{Garvin}, J.~B., {Getty}, S.~A., {Arney}, G.~N., {et~al.} 2022, \psj, 3, 117,
  \dodoi{10.3847/PSJ/ac63c2}

\bibitem[{{Goldreich} \& {Peale}(1966)}]{goldreich1966f}
{Goldreich}, P., \& {Peale}, S. 1966, \aj, 71, 425, \dodoi{10.1086/109947}

\bibitem[{{Goldreich} \& {Soter}(1966)}]{goldreich1966a}
{Goldreich}, P., \& {Soter}, S. 1966, \icarus, 5, 375,
  \dodoi{10.1016/0019-1035(66)90051-0}

\bibitem[{{Green} {et~al.}(2019){Green}, {Way}, \& {Barnes}}]{green2019}
{Green}, J.~A.~M., {Way}, M.~J., \& {Barnes}, R. 2019, \apjl, 876, L22,
  \dodoi{10.3847/2041-8213/ab133b}

\bibitem[{{Holman} \& {Wiegert}(1999)}]{holman1999}
{Holman}, M.~J., \& {Wiegert}, P.~A. 1999, \aj, 117, 621,
  \dodoi{10.1086/300695}

\bibitem[{{Huang} {et~al.}(2025){Huang}, {Ormel}, {Portegies Zwart}, {Kokubo},
  \& {Yi}}]{huang2025c}
{Huang}, S., {Ormel}, C.~W., {Portegies Zwart}, S., {Kokubo}, E., \& {Yi}, T.
  2025, \apj, 988, 137, \dodoi{10.3847/1538-4357/ade25a}

\bibitem[{{Hut}(1981)}]{hut1981c}
{Hut}, P. 1981, \aap, 99, 126

\bibitem[{{Jacobson} {et~al.}(2017){Jacobson}, {Rubie}, {Hernlund},
  {Morbidelli}, \& {Nakajima}}]{jacobson2017}
{Jacobson}, S.~A., {Rubie}, D.~C., {Hernlund}, J., {Morbidelli}, A., \&
  {Nakajima}, M. 2017, Earth and Planetary Science Letters, 474, 375,
  \dodoi{10.1016/j.epsl.2017.06.023}

\bibitem[{{Johansen} {et~al.}(2021){Johansen}, {Ronnet}, {Bizzarro},
  {Schiller}, {Lambrechts}, {Nordlund}, \& {Lammer}}]{johansen2021}
{Johansen}, A., {Ronnet}, T., {Bizzarro}, M., {et~al.} 2021, Science Advances,
  7, eabc0444, \dodoi{10.1126/sciadv.abc0444}

\bibitem[{{Kane}(2017)}]{kane2017c}
{Kane}, S.~R. 2017, \apjl, 839, L19, \dodoi{10.3847/2041-8213/aa6bf2}

\bibitem[{{Kane}(2022)}]{kane2022b}
---. 2022, Nature Astronomy, 6, 420, \dodoi{10.1038/s41550-022-01626-x}

\bibitem[{{Kane} \& {Byrne}(2024)}]{kane2024b}
{Kane}, S.~R., \& {Byrne}, P.~K. 2024, Nature Astronomy, 8, 417,
  \dodoi{10.1038/s41550-024-02228-5}

\bibitem[{{Kane} {et~al.}(2014){Kane}, {Kopparapu}, \&
  {Domagal-Goldman}}]{kane2014e}
{Kane}, S.~R., {Kopparapu}, R.~K., \& {Domagal-Goldman}, S.~D. 2014, \apjl,
  794, L5, \dodoi{10.1088/2041-8205/794/1/L5}

\bibitem[{{Kane} {et~al.}(2019){Kane}, {Arney}, {Crisp}, {Domagal-Goldman},
  {Glaze}, {Goldblatt}, {Grinspoon}, {Head}, {Lenardic}, {Unterborn}, {Way}, \&
  {Zahnle}}]{kane2019d}
{Kane}, S.~R., {Arney}, G., {Crisp}, D., {et~al.} 2019, Journal of Geophysical
  Research (Planets), 124, 2015, \dodoi{10.1029/2019JE005939}

\bibitem[{{Konopliv} \& {Yoder}(1996)}]{konopliv1996}
{Konopliv}, A.~S., \& {Yoder}, C.~F. 1996, \grl, 23, 1857,
  \dodoi{10.1029/96GL01589}

\bibitem[{{Korenaga}(2023)}]{korenaga2023a}
{Korenaga}, J. 2023, \icarus, 400, 115564, \dodoi{10.1016/j.icarus.2023.115564}

\bibitem[{{Laskar} {et~al.}(1993){Laskar}, {Joutel}, \&
  {Robutel}}]{laskar1993b}
{Laskar}, J., {Joutel}, F., \& {Robutel}, P. 1993, \nat, 361, 615,
  \dodoi{10.1038/361615a0}

\bibitem[{{Leconte} {et~al.}(2010){Leconte}, {Chabrier}, {Baraffe}, \&
  {Levrard}}]{leconte2010a}
{Leconte}, J., {Chabrier}, G., {Baraffe}, I., \& {Levrard}, B. 2010, \aap, 516,
  A64, \dodoi{10.1051/0004-6361/201014337}

\bibitem[{{Leconte} {et~al.}(2015){Leconte}, {Wu}, {Menou}, \&
  {Murray}}]{leconte2015a}
{Leconte}, J., {Wu}, H., {Menou}, K., \& {Murray}, N. 2015, Science, 347, 632,
  \dodoi{10.1126/science.1258686}

\bibitem[{{Levrard} {et~al.}(2007){Levrard}, {Correia}, {Chabrier}, {Baraffe},
  {Selsis}, \& {Laskar}}]{levrard2007a}
{Levrard}, B., {Correia}, A.~C.~M., {Chabrier}, G., {et~al.} 2007, \aap, 462,
  L5, \dodoi{10.1051/0004-6361:20066487}

\bibitem[{{Lissauer} {et~al.}(2012){Lissauer}, {Barnes}, \&
  {Chambers}}]{lissauer2012a}
{Lissauer}, J.~J., {Barnes}, J.~W., \& {Chambers}, J.~E. 2012, \icarus, 217,
  77, \dodoi{10.1016/j.icarus.2011.10.013}

\bibitem[{{Makarov} \& {Efroimsky}(2025)}]{makarov2025d}
{Makarov}, V.~V., \& {Efroimsky}, M. 2025, Universe, 11, 309,
  \dodoi{10.3390/universe11090309}

\bibitem[{{Makarov} \& {Goldin}(2024)}]{makarov2024a}
{Makarov}, V.~V., \& {Goldin}, A. 2024, Universe, 10, 15,
  \dodoi{10.3390/universe10010015}

\bibitem[{{Margot} {et~al.}(2021){Margot}, {Campbell}, {Giorgini}, {Jao},
  {Snedeker}, {Ghigo}, \& {Bonsall}}]{margot2021c}
{Margot}, J.-L., {Campbell}, D.~B., {Giorgini}, J.~D., {et~al.} 2021, Nature
  Astronomy, \dodoi{10.1038/s41550-021-01339-7}

\bibitem[{{Murray} \& {Dermott}(1999)}]{murray1999a}
{Murray}, C.~D., \& {Dermott}, S.~F. 1999, {Solar System Dynamics} ({Cambridge
  University Press}), \dodoi{10.1017/CBO9781139174817}

\bibitem[{{Nesvorn{\'y}} {et~al.}(2025){Nesvorn{\'y}}, {Morbidelli}, {Bottke},
  {Deienno}, \& {Goldberg}}]{nesvorny2025b}
{Nesvorn{\'y}}, D., {Morbidelli}, A., {Bottke}, W.~F., {Deienno}, R., \&
  {Goldberg}, M. 2025, \aj, 170, 180, \dodoi{10.3847/1538-3881/adf20a}

\bibitem[{{Nimmo} {et~al.}(2024){Nimmo}, {Kleine}, \&
  {Morbidelli}}]{nimmo2024c}
{Nimmo}, F., {Kleine}, T., \& {Morbidelli}, A. 2024, \nat, 636, 598,
  \dodoi{10.1038/s41586-024-08231-0}

\bibitem[{{Ostberg} {et~al.}(2023){Ostberg}, {Kane}, {Li}, {Schwieterman},
  {Hill}, {Bott}, {Dalba}, {Fetherolf}, {Head}, \& {Unterborn}}]{ostberg2023a}
{Ostberg}, C., {Kane}, S.~R., {Li}, Z., {et~al.} 2023, \aj, 165, 168,
  \dodoi{10.3847/1538-3881/acbfaf}

\bibitem[{{Peale}(1999)}]{peale1999a}
{Peale}, S.~J. 1999, \araa, 37, 533, \dodoi{10.1146/annurev.astro.37.1.533}

\bibitem[{{Renaud} \& {Henning}(2018)}]{renaud2018}
{Renaud}, J.~P., \& {Henning}, W.~G. 2018, \apj, 857, 98,
  \dodoi{10.3847/1538-4357/aab784}

\bibitem[{{Stevenson}(2003)}]{stevenson2003}
{Stevenson}, D.~J. 2003, Earth and Planetary Science Letters, 208, 1,
  \dodoi{10.1016/S0012-821X(02)01126-3}

\bibitem[{{Touma} \& {Wisdom}(1994)}]{touma1994b}
{Touma}, J., \& {Wisdom}, J. 1994, \aj, 108, 1943, \dodoi{10.1086/117209}

\bibitem[{{Ward} \& {Reid}(1973)}]{ward1973a}
{Ward}, W.~R., \& {Reid}, M.~J. 1973, \mnras, 164, 21,
  \dodoi{10.1093/mnras/164.1.21}

\bibitem[{{Way} \& {Del Genio}(2020)}]{way2020}
{Way}, M.~J., \& {Del Genio}, A.~D. 2020, Journal of Geophysical Research
  (Planets), 125, e06276, \dodoi{10.1029/2019JE006276}

\end{thebibliography}


\end{document}